\documentclass[11pt,letterpaper]{article}

\usepackage[english]{babel}
\usepackage{soul} 
\usepackage[normalem]{ulem}  
\usepackage{setspace}  
\usepackage{xcolor}  
\usepackage{color}  
\usepackage{setspace}  
\usepackage[compact,md]{titlesec}  
\usepackage{textcomp}  
\usepackage[version=4]{mhchem}  
\usepackage{siunitx}  

\usepackage[utf8]{inputenc}  
\usepackage{authblk}  
\usepackage{blindtext}  
\usepackage{xurl}
\usepackage{etoolbox}  
\usepackage{tcolorbox}  
\usepackage[export]{adjustbox}  
\usepackage{subfiles}  

\date{}

\makeatletter
\def\blfootnote{\gdef\@thefnmark{}\@footnotetext}
\makeatother

\usepackage{graphicx}  
\usepackage{pdfpages}  
\usepackage{pgfgantt}  
\usepackage{caption}  
\usepackage{subcaption}  
\usepackage{rotating}  
\usepackage{wrapfig}  
\usepackage{tabularx}  
\usepackage{multirow}  
\usepackage{array}  
\usepackage{longtable}  
\usepackage{makecell}  
\usepackage{colortbl}  

\usepackage{amsmath}  
\usepackage{amsfonts}  
\usepackage{amssymb}  
\usepackage{amsthm}   
\usepackage{esdiff}  
\usepackage{siunitx}  
\usepackage{cancel}  
\usepackage{bm}  

\usepackage{doi}  
\usepackage{hypernat} 
\usepackage{pdfsync}  
\usepackage[sorting=none,backend=biber,maxnames=50,citestyle=numeric-comp]{biblatex}

\usepackage{algorithm}  
\usepackage{algpseudocode}  
\usepackage{listings}  

\titlespacing{\section}{0pt}{1.5ex}{1.5ex}
\titlespacing{\subsection}{0pt}{1.5ex}{1.5ex}
\titlespacing{\subsubsection}{0pt}{1.ex}{1.ex}

\titleformat{\section}[block]
{\center\normalfont\scshape}  
{\textcolor{ucsd_blue}{\thesection}}  
{0.5em}  
{\color{ucsd_blue}}  
[]  

\titleformat{\subsection}[block]
{\normalfont}  
{\textcolor{ucsd_blue}{\thesubsection}}  
{0.5em}  
{\color{ucsd_blue}}  
[]  

\titleformat{\subsubsection}[block]
{\normalfont}  
{\textcolor{ucsd_blue}{\thesubsubsection}}  
{0.5em}  
{\color{ucsd_blue}}  
[]  

\newcommand\preprintheaderandtitle[3]{
    \begin{tcolorbox}[boxrule=0.75pt, arc=2pt, coltext=ucsd_blue, colback=white,colframe=ucsd_gold, fontupper=\rmfamily]
        \footnotesize
        \setstretch{1.25}
        {\textcolor{ucsd_gray}
        {This is the preprint version of the following article:}}

        #1

        \vspace{1mm}

        \textcolor{ucsd_gray}{Published article:} \;
        \url{#2}

    \end{tcolorbox}
    
    \vspace{-6mm}

    {\let\newpage\relax\maketitle}

    \vspace{-6mm}
}
\definecolor{ucsd_blue}{RGB}{24, 43, 73}
\definecolor{ucsd_gold}{RGB}{198, 146, 20}
\definecolor{ucsd_blue_light}{RGB}{0, 98, 155}
\definecolor{ucsd_gold_light}{RGB}{255, 205, 0}
\definecolor{ucsd_gray}{RGB}{116, 118, 120}
\newcommand\p\partial  

\newcommand\f\frac  

\begin{document}

\title{
    \textbf{
        Large-scale multidisciplinary design optimization of the NASA lift-plus-cruise concept using a novel aircraft design framework
    }
    \blfootnote{
        This paper was originally presented at the VFS Autonomous VTOL Technical Meeting, Mesa, Arizona, USA, Jan. 24-26, 2023.
    }
}

\author{
    Marius L. Ruh\footnote{PhD Student, Department of Mechanical and Aerospace Engineering}, \enspace
    Darshan Sarojini\footnote{Postdoctoral fellow, Department of Mechanical and Aerospace Engineering},  \enspace
    Andrew Fletcher\footnote{PhD Student, Department of Mechanical and Aerospace Engineering}}
\affil{University of California San Diego, La Jolla, CA, 92093}

\author{
    \\ Isaac Asher \footnote{Staff engineer}
}
\affil{Aurora Flight Sciences, Manassas, VA, 20110}

\author{
     John T. Hwang\footnote{Assistant professor
}}
\affil{University of California San Diego, La Jolla, CA, 92093}

\renewcommand\Affilfont{\itshape\small}

\begin{refsection}

    \preprintheaderandtitle
    {
        Marius L Ruh, Darshan Sarojini, Andrew Fletcher, Isaac Ahser, and John T. Hwang.  Large-scale multidisciplinary design optimization of the NASA lift-plus-cruise concept using a novel aircraft design framework.
    }
    {https://vtol.org/store/product/largescale-multidisciplinary-design-optimization-of-the-nasa-liftpluscruise-concept-using-a-novel-aircraft-design-framework-17776.cfm}

\abstract{
The conceptual design of eVTOL aircraft is a high-dimensional optimization problem that involves large numbers of continuous design parameters.
Therefore, eVTOL design methods would benefit from numerical optimization algorithms capable of systematically searching these high-dimensional parameter spaces, using comprehensive and multidisciplinary models of the aircraft.
By leveraging recent progress in sensitivity analysis methods, a computational framework called the Comprehensive Aircraft high-Dimensional DEsign Environment (CADDEE) has been developed for large-scale multidisciplinary design optimization (MDO) of electric air taxis.
CADDEE uses a geometry-centric approach that propagates geometry changes in a differentiable manner to meshes for physics-based models of arbitrary fidelity level. 
This paper demonstrates the capabilities of this new aircraft design tool, by presenting large-scale MDO results for NASA’s Lift+Cruise eVTOL concept. 
MDO with over 100 design variables, 17 constraints, and low-fidelity predictive models for key disciplines is demonstrated with an optimization time of less than one hour with a desktop computer. 
The results show a reduction in gross weight of 11.4\% and suggest that CADDEE can be valuable in the conceptual design and optimization of eVTOL aircraft.
}

    \section{Introduction}

\label{sec:intro}


A push toward electrification in aviation has contributed to the development of novel and efficient aircraft with vertical takeoff and landing (eVTOL) capabilities.
The envisioned potential of such aircraft in areas such as urban and regional air mobility (UAM/RAM), delivery drones, military applications, and others has 
led to active research in conceptual design for this new class of vehicles. 
Established conceptual design approaches~\cite{raymer2012aircraft, roskam2003airplane, torenbeek2013synthesis} and software frameworks such as NASA's Flight Optimization System (FLOPS)~\cite{mccullers1984aircraft}, NASA's Design and Analysis of Rotorcraf (NDARC)~\cite{johnson2010ndarc}, or DLR's Common Parametric Aircraft Configuration Scheme (CPACS)~\cite{nagel2012communication} are successful at rapid design and optimization of conventional air-and rotorcraft configurations.
This is due to an abundance of historical data that can be used to train regression models, which can quickly explore and optimize in the known design space of conventional aircraft.

The design space of eVTOL vehicles, for which over 700 different concepts~\cite{eVTOLconcepts} have been proposed, is not yet fully explored.
With a lack of established and proven designs, traditional conceptual design approaches have limited success and may produce unreliable results. 
To address this limitation, NASA has developed a successor to FLOPS, called Layered and Extensible Aircraft Performance System (LEAPS)~\cite{welstead2018overview} that enables modeling of advanced aircraft concepts, with a focus on electric and hybrid-electric propulsion systems.
Another recently introduced design framework is called SUAVE~\cite{lukaczyk2015suave}, which focuses on unconventional aircraft configurations, augmenting existing regressions with low-fidelity physics-based models. 
An energy-based sizing framework called PEACE~\cite{chakraborty2021generalized} was developed for new concepts and unconventional propulsion architectures and describes design disciplines in a functional form that allows for a combination of table data, surrogates, and physics-based models.
In a survey, Johnson and Silva~\cite{johnson2022nasa} provide an overview of the challenges associated with the conceptual design of UAM vehicles and the limitations of applying traditional aircraft design frameworks. 

Other significant research that can benefit the design of eVTOL vehicles is summarized as follows: sizing of novel aircraft concepts \cite{bryson2016multidisciplinary,hascaryo2020configuration}, geometry-centric design \cite{lukaczyk2015suave,hwang2012geomach}, unconventional propulsion architectures \cite{de2020range,finger2020comparison}, dynamic loads analysis \cite{sarojini2021dynamic}, physics-based structural sizing~\cite{solano2021parametric}, and safety, reliability and robust design \cite{chaudhuri2022certifiable,bendarkar2022off}.


While recent research has provided new methodologies for eVTOL aircraft design and analysis, there is a need for a general and comprehensive design framework that can rapidly explore a high-dimensional design space while considering all relevant disciplines.
This work presents results from an ongoing research effort that aims to address this need.
This ongoing effort centers around a new design framework called the Comprehensive Aircraft high-Dimensional DEsign Environment (CADDEE). 
CADDEE is unique in relation to the aircraft design frameworks surveyed above 
because it enables large-scale (high-dimensional) multidisciplinary design optimization (MDO) that relies on gradient-based optimizers to efficiently scale up to hundreds of design variables.
CADDEE's critical feature is that it is implemented in a recently developed algebraic modeling language that automates derivative computation. 
CADDEE also adopts a geometry-centric approach to MDO that allows for changes in the geometry to be propagated using smooth mappings to physics-based models.

\begin{figure}[!hbt] \begin{center}
\includegraphics[scale=1]{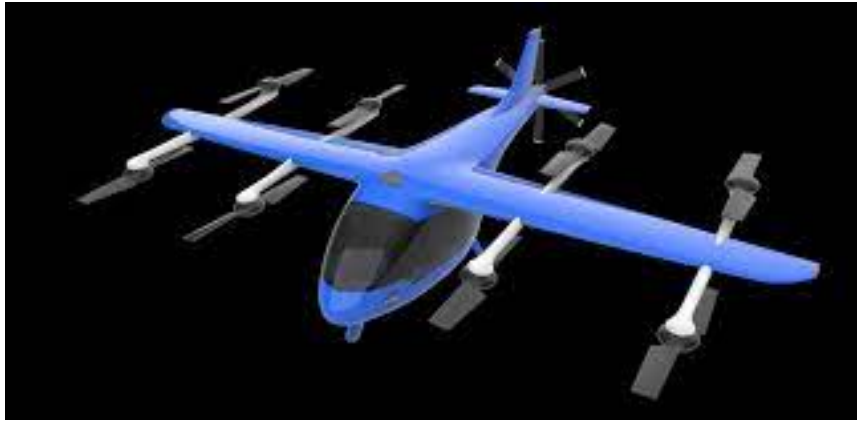}
\caption{NASA Lift+Cruise concept vehicle}\label{fig:NASA LPC concept}
\end{center}\end{figure}

As a demonstration of CADDEE's capabilities, this paper presents results from a set of large-scale MDO studies of a NASA Lift+Cruise concept vehicle (Fig.~\ref{fig:NASA LPC concept}).
These studies explore the design space consisting of over 100 design variables for different optimization objectives and constraints, verifying existing limiting factors in the design of eVTOL vehicles, such as rotor noise and battery technology. 
Results show the potential of a capability for large-scale, comprehensive analysis and design of eVTOL vehicles early in the conceptual design phase.

In the remainder of this paper, the CADDEE framework is presented first before giving an overview of the discipline models used in the analyses. 
Then, results from aeroacoustic and large-scale MDO studies are shown, including Pareto fronts and parameter sweeps.

    \section{CADDEE Design Framework}
CADDEE is a software library for constructing multidisciplinary models of aircraft (and similar systems) in a manner that facilitates  analysis and design using large-scale MDO. 
In this section, the features and general architecture of this framework are summarized. 
A paper from the same research effort~\cite{sarojini2023tc1} is presented at the AIAA Scitech 2023 Forum and explains the framework in greater detail. 
It should be noted that CADDEE is still under active development and not all of the described features are fully mature.

\subsection{Design Parameterization}
\begin{figure}[!hbt] \begin{center}
\includegraphics[scale=1]{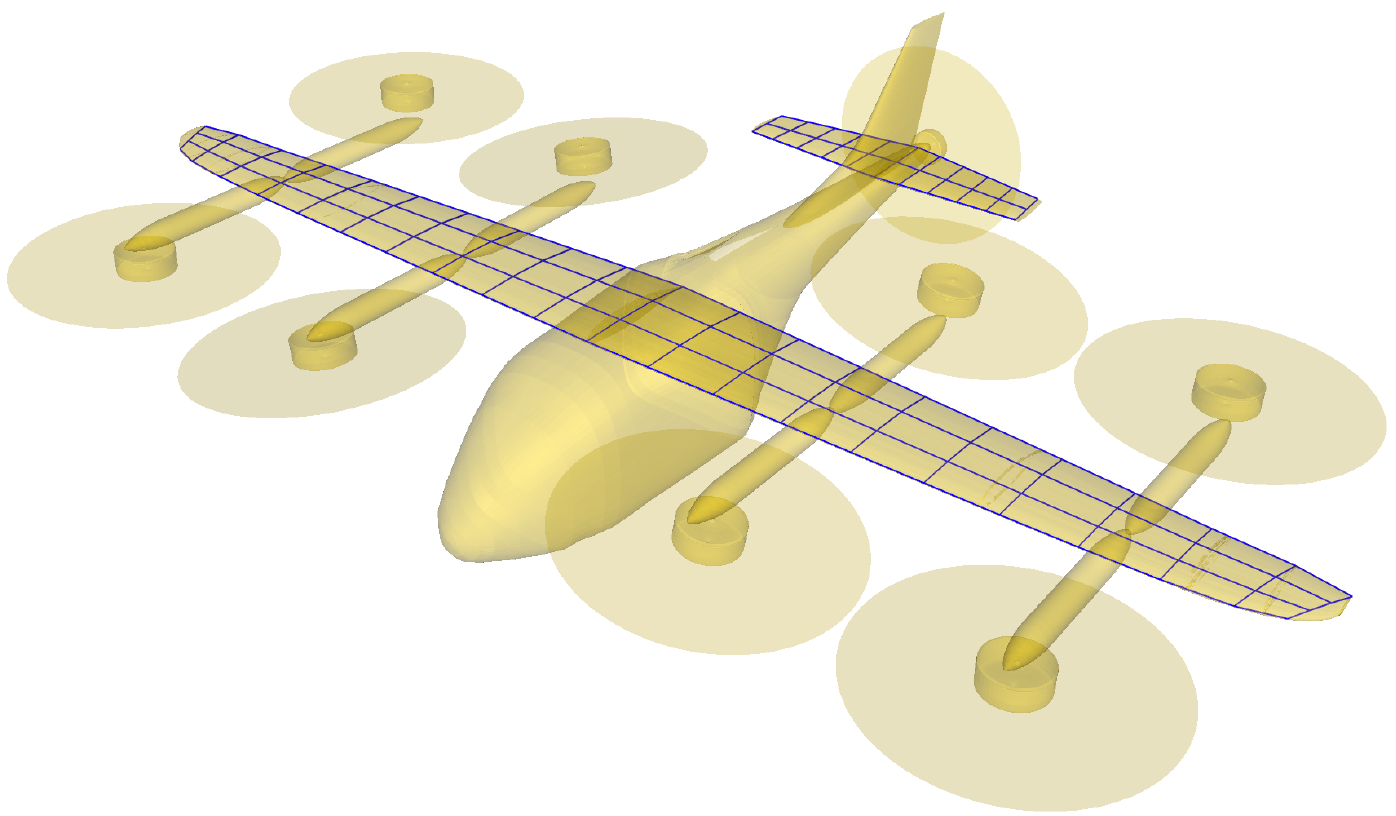}
\caption{Illustration of the use of mapped arrays to create a camber surface mesh for the wing and horizontal tail.}\label{fig:camber mesh}
\end{center}\end{figure}

\begin{figure}[!hbt] \begin{center}
\includegraphics[scale=1]{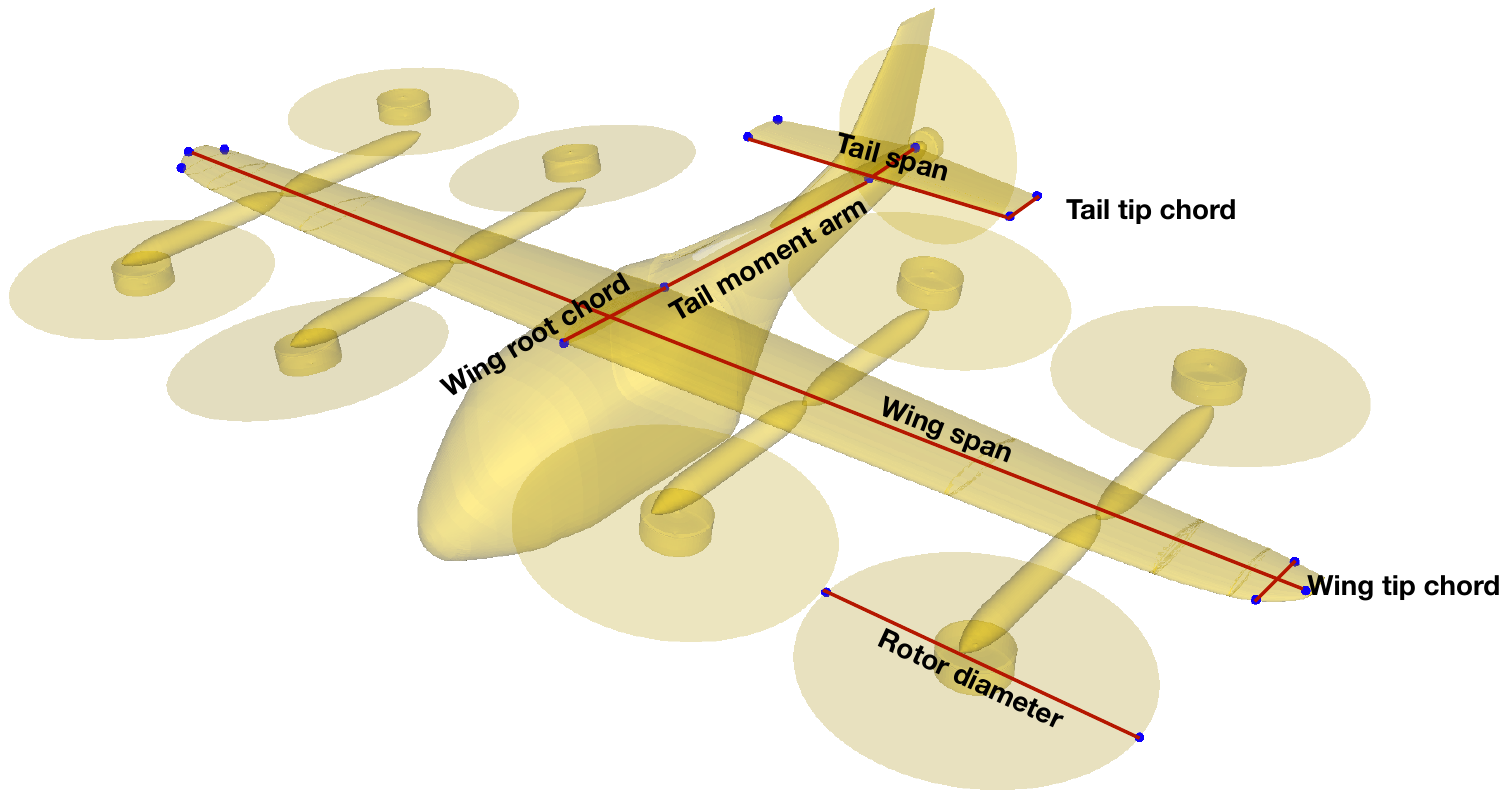}
\caption{Illustration of the use of mapped arrays to compute high-level geometric quantities.}\label{fig:geom outputs}
\end{center}\end{figure}

The design is parameterized using a high-fidelity representation of the outer mold line (OML). 
B-spline surfaces are used due to their analytic form and their ability to represent complex geometries with a small number of control points. 
The B-spline representation is then used to establish relationships between any (external) set of points and the central geometry via so-called \textit{mapped arrays}. 
A mapped array is a set of points resulting from the evaluation of a parametric map, which is determined by projecting points onto the OML.
For example, mapped arrays are used to extract meshes, determine geometric variables such as wing area, rotation axes, and parameters for bulk actuation such as a tilting rotor or wing. 
By defining these mapped arrays in terms of the B-spline surfaces, any changes to the design geometry can be efficiently propagated through a simple matrix multiplication. 
This is especially useful during optimization, where changes in geometric variables such as wing span or the location of a rotor need to be accurately represented such that discipline models like aerodynamics and propulsion take into account changes in the geometry. 
The application of mapped arrays is shown in Fig.~\ref{fig:camber mesh} where a camber surface mesh for the wing and horizontal tail is generated using mapped arrays and is superimposed with the B-spline geometry. 
In Fig.~\ref{fig:geom outputs}, high-level sizing parameters such as wing span, rotor radius and others are determined from relative distances using mapped arrays, which are shown by the blue points.

\subsection{Automated Derivative Computation}
Another feature of CADDEE is that the backend is implemented in a new, domain-embedded programming language called Computational System Design Language (CSDL)~\cite{gandarillas2022novel} that is well-suited for  constructing multidisciplinary numerical engineering models that may also be coupled.
CSDL fully automates adjoint-based sensitivity analysis of any mathematical operation, which allows for the efficient computation of derivatives of objective functions and constraints with respect to design variables. 
In combination with gradient-based numerical optimization algorithms such as SNOPT~\cite{gill2005snopt}, automated derivative computation allows CADDEE to perform large-scale MDO with up to hundreds of design variables, as demonstrated in this paper.

\subsection{Comprehensive Modeling}
CADDEE is able to model all relevant aircraft design disciplines for a full mission, consisting of multiple mission segments. 
Segments may contain different discipline models of varying fidelity. 
Models of the same kind are vectorized across misison segments for improved computational efficiency. 
In general, mission segments may be of three different kinds. 
First, trim-state segments describe on-design, steady operational conditions such as cruise or hover. 
These are the simplest analysis cases. 
Second, transient segments require time integration schemes, e.g., to model dynamic transition or a gust response. 
Such conditions are more computationally expensive. 
The third kind of mission segment models off-design conditions such as flutter or aeroelastic divergence. 

\begin{figure}[ht] \begin{center}
\includegraphics[scale=1]{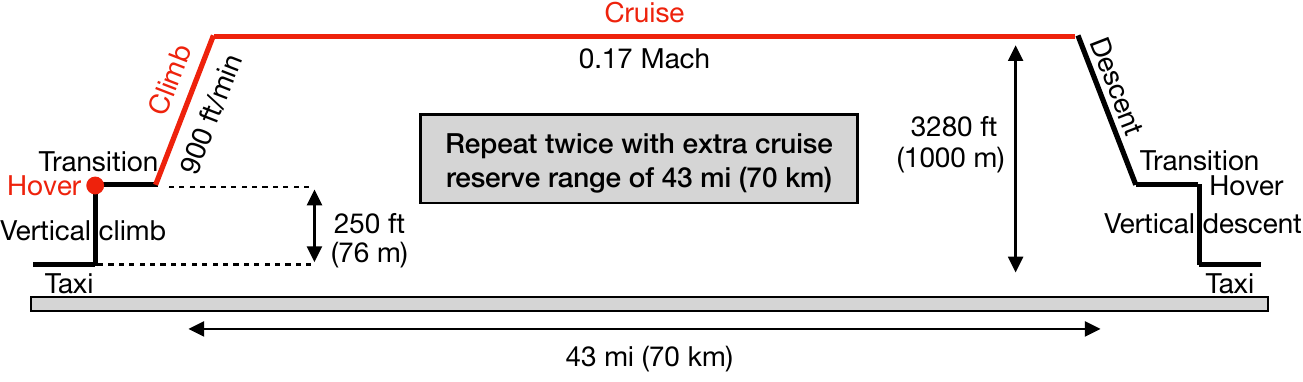}
\caption{Example of a mission profile that can be constructed using CADDEE.}\label{fig:misison profile}
\end{center}
\end{figure}

For the results presented in this paper, a simple mission is chosen, consisting of three steady (trim-state) segments, hover, climb, and cruise. 
These segments are a subset of a sample mission proposed in an eVTOL design study by NASA~\cite{silva2018vtol}, which also includes taxi, dynamic transition, and descent. 
A modified version of the mission profile is shown in Fig.~\ref{fig:misison profile}, where the modeled segments are indicated in red.
The hover segment is at an altitude of 250 ft and lasts for 90 seconds. 
The shown mission is repeated twice and has a 43 mile cruise reserve, that is included in the model.

\subsection{Framework Architecture}
\begin{figure*}[!hbt] \begin{center}
\includegraphics[scale=0.7]{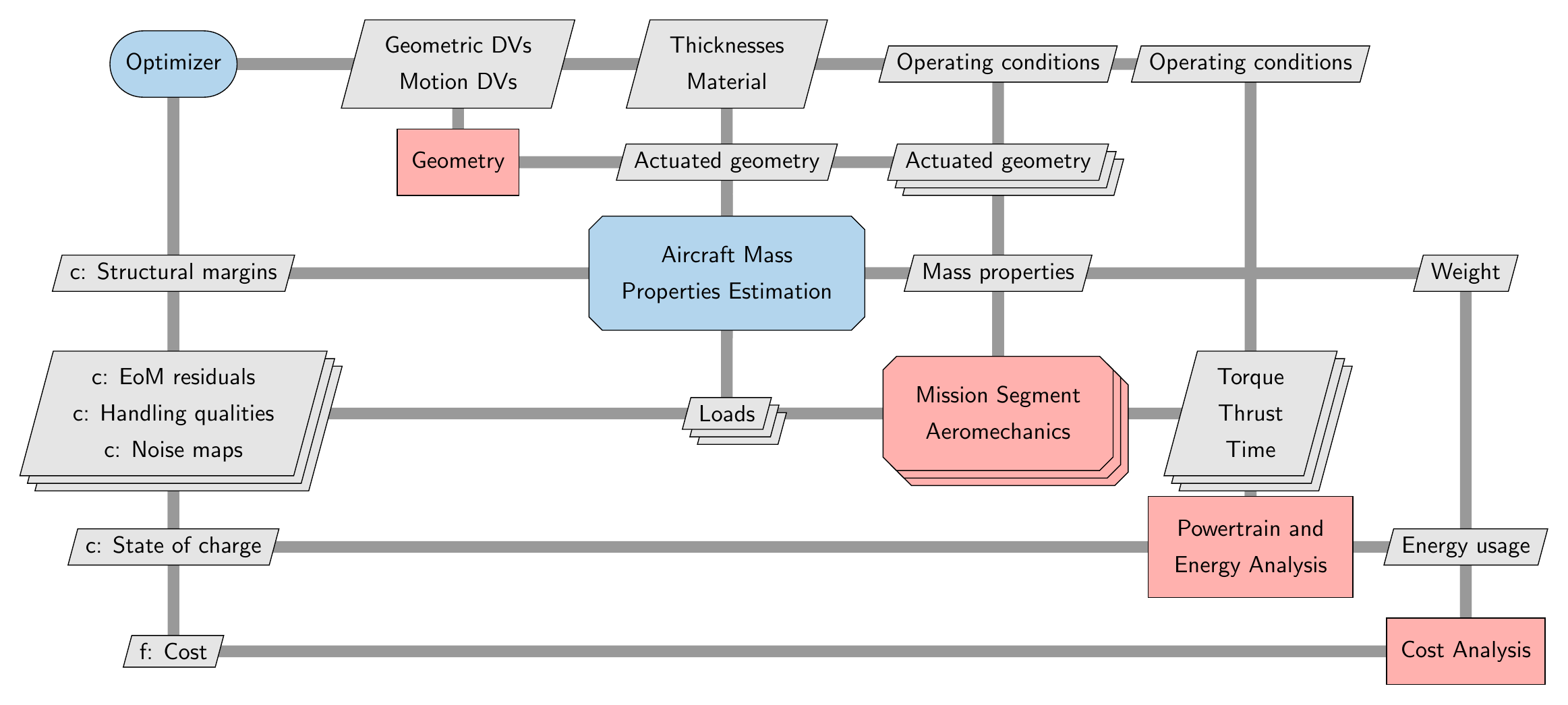}
\caption{Design structure matrix of the CADDEE framework architecture}\label{fig:caddee architecture}
\end{center}\end{figure*}

Figure~\ref{fig:caddee architecture} illustrates the general architecture of the CADDEE framework in the form of a design structure matrix as it applies to the test case presented in this paper. 
It is shown how disciplines are connected, what constraints are imposed (indicated by \textit{c}) and what variables are optimized (top row).  
The cost analysis outputs the objective function (indicated by \textit{f}); 
however, this can be any other quantity of interest such as gross weight, range, noise etc. 
The geometry model takes in geometric design variables (e.g., wing span, wing area) and motion design variables (e.g., tilt rotation) and outputs an actuated geometry that contains meshes and other geometric quantities needed by the discipline models. 
The actuated geometry, along with information about material and thicknesses, is passed into the mass properties model, which outputs aircraft-level mass properties from individual subsystem components. 

The next block, loosely labeled ``Aeromechanics", contains the bulk of the analysis, vectorized across all mission segments, which is indicated by the stacked shapes. 
Models that are included in the aeromechanics block are aerodynamics, propulsion, structural analysis, stability and control, as well as acoustics.
In general, coupling between disciplines may arise, depending on the depth of the analysis. 
Depending on the model and its fidelity, inputs to the aeromechanics block are the previously computed mass properties, actuated geometry, and the operating conditions of the aircraft (e.g., Mach number, pitch angle, rotor rotational speed).
Some outputs of the aeromechanics block can be constraints, depending on the optimization problem. 
Such constraints may concern the trim state of the aircraft (EoM residuals), stability and control (handling qualities), or acoustics (noise maps). 
Other aeromechanics outputs are passed into the last major analysis block, containing the powertrain and energy analysis. 
Depending on the architecture of the powertrain (e.g., for eVTOL aircraft), inputs are the torque required by the rotors and the total time spent in each mission segment. 
Theses are required as inputs by the motor analysis to compute a power profile for the mission that is used by the battery model to estimate the final state-of-charge, which may be a constraint during optimization. 
For future versions of CADDEE, intermediate components such as inverters and gearboxes that connect motors, batteries and rotors, may also be modeled. 
Lastly, the output of the powertrain and energy analysis is the total energy used, which may be passed into the cost analysis along with the aircraft gross weight, if such a model is implemented. 

    \section{Discipline Models}
Most models used in this paper are low-fidelity, physics-based or empirical models. 
Models are tailored for eVTOL design and are developed in a robust and computationally efficient way, suited for large-scale MDO. 

\subsection{Structural Sizing and Weight Estimation}
\begin{figure}[!hbt] \begin{center}
\includegraphics[scale=1.2]{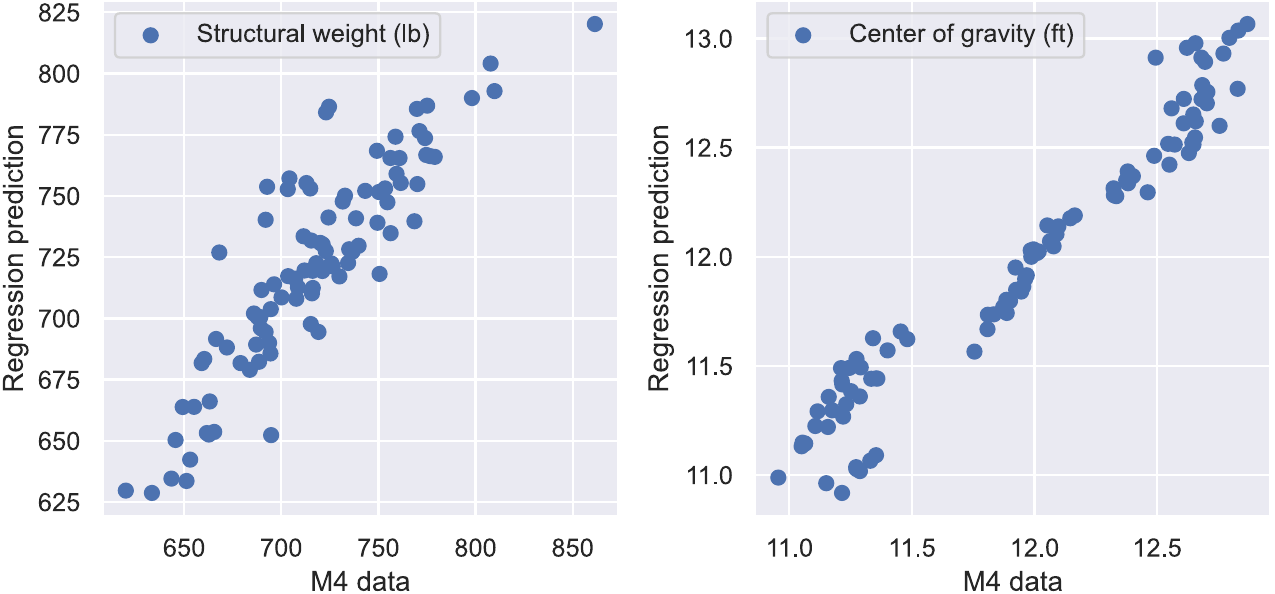}
\caption{Verification of sizing surrogate model}\label{fig:M4 data}
\end{center}\end{figure}

For traditional aircraft, empirical models based on historic data can be used to estimate the aircraft weight.
However, this is not the case for UAM aircraft.
To create a new empirical model, physics-based structural sizing is performed via a software tool called M4 Structures Studio, which has been used to size a number of UAM concepts~\cite{winter2020structural, winter2021conceptual, winter2021crashworthiness}.
The software parameterizes the aircraft to create a finite element model for NASTRAN.
In this work, the Lift+Cruise concept is parameterized in terms of wing area, wing aspect ratio, fuselage length, cruise speed, battery weight, horizontal tail area, and vertical tail area. 
The design space spanned by these seven variables is sampled using the Latin Hypercube sampling technique with upper and lower bounds of $\pm$10\% based on the baseline configuration. 
Each combination of design inputs is used to create a new finite element model and the structural weight and other mass properties (e.g., center of gravity) are determined using NASTRAN's solution 200. 
A total of 96 converged optimizations provide the set of data based on which multi-variate linear regression models are trained, to predict aircraft-level mass properties instantaneously. 
Scaling factors are added such that the weight predictions match those obtained by NDARC~\cite{johnson2017ndarc} for the baseline Lift+Cruise configuration.
An average $\textrm{r}^2$ score of 0.94 is achieved by the surrogate models and two sample scatter plots are shown in Fig.~\ref{fig:M4 data} of the training data versus the corresponding prediction by the regression model. 
The quantities shown are the combined structural weight of the booms, fuselage, and wing as well as the x-location of the combined center of gravity.

\subsection{Aerodynamic Model}
The aerodynamic model used is the vortex-lattice method (VLM), which is based on potential flow theory.  
The model assumptions are that the flow is steady, inviscid, incompressible, and irrotational. 
In addition, the lifting surface is assumed to be thin and at a small angle of attack. 
Model inputs are flow velocity, air density, and a camber surface mesh for each lifting surface that is modeled.
In this study, the wing and horizontal tail are considered as lifting surfaces. 
Outputs of the VLM model are the traction forces on each panel, which are computed from the Kutta--Jukowski theorem. 
The model is validated against a low-fidelity aerostructural analysis tool called OpenAerostruct~\cite{jasa2018open}.
Lastly, the drag due to other components of the aircraft such as the fuselage, rotor hubs, or landing gear is computed using drag area estimates provided by Silva et al.~\cite{silva2018vtol} for the Lift+Cruise concept.

\subsection{Propulsion Model}
Blade-element momentum (BEM) theory is used to model the rotor aerodynamics. 
The theory combines momentum and blade element theory to solve a system of coupled equations to predict the sectional thrust and torque distribution. 
In the solution process, a residual function needs to be solved to predict the sectional inflow angle. 
The implementation used in this paper~\cite{ruh2021robust} guarantees a solution of the residual function for all reasonable operating conditions. 
Model inputs to BEM are the inflow velocity, rotational speed, blade geometry given by chord and twist, and an airfoil polar to predict the sectional lift and drag coefficient. 
In this work, an airfoil surrogate model is trained~\cite{bouhlel2019python} based on data obtained from XFOIL~\cite{drela1989xfoil} for a NACA 4412 airfoil.
The airfoil model predicts the sectional lift and drag coefficient as a function of angle of attack and Reynolds number.
The same airfoil is used for the entire rotor span.
Lastly, the Prandtl tip-loss correction model is used to acount for losses near the rotor tip.

\subsection{Stability Analysis}
In this paper, the stability analysis is limited to trimming the aircraft in each mission segment, such that the net forces and moments are zero. 
A set of rigid-body equations of motion, expressed in the body-fixed frame, is used to derive a residual, expressing the linear and angular accelerations due to the forces and moments acting on the aircraft.
During optimization, this residual is optimized such that it approaches zero. 
The equations of motion assume that the surface of the Earth is flat, rotational effect are negligible, and the atmosphere is at rest.
In addition, the moments of inertia are treated as constant. 
Inputs to the equations of motion are aerodynamic and inertial forces, linear and angular accelerations, and system mass properties.
Future versions of CADDEE will also compute handling qualities, following guidelines based on regulations~\cite{moorhouse1980us, ads33}.

\subsection{Acoustic Model}
Rotor noise is comprised of many different sources. 
In this paper, tonal and broadband noise are considered. 
Tonal noise, also known as rotational noise, is low-frequency sound produced by periodic sound pressure disturbances due to rotation.
This noise source can be further divided into loading and thickness noise, the latter of which captures noise due to the blade geometry. 
In this paper, a frequency-domain method developed by Barry and Magliozzi~\cite{Barry:1971:AFAPL} is used to predict tonal noise, which is suited for axial flight and hover. 
A limitation of this model is that it significantly under predicts the sound pressure level at the rotor center directly below the rotor disk, which is shown in more detail later. 
Broadband noise, also referred to as vortex noise, is high-frequency sound that is generated by random fluctuations of the forces on the blade that arise due to turbulence.
The prediction of broadband noise is more difficult than tonal noise due to its stochastic nature and multiple sources. 
For this study, an empirical broadband noise model is used,  developed by Schlegel et al.~\cite{Schlegel:1966:USAAML} based on helicopter experiments.
The tonal and broadband noise models used here have been validated by Herniczek et al.~\cite{Herniczek:2017:AIAA, Herniczek:2019:AIAA} and Gill et al.~\cite{Gill:2023:AIAA}.
Both sources of noise are combined to determine the total sound pressure level of a rotor, expressed in A-weighted decibel (dB-A). 
A-weighting captures human annoyance by weighing frequencies in the human hearing spectrum more heavily. 

\subsection{Motor Sizing and Analysis}
For large-scale analyses, motor sizing and analysis are typically done using surrogate models with simplifying assumptions such as constant motor efficiency and empirical weight estimation based on torque-to-weight and power-to-weight ratios. 
In this study, a low-fidelity physics-based model~\cite{cheng2023differentiable} is used for sizing and analysis, based on an equivalent circuit model and a novel model for motor control.
This model considers both maximum torque per ampere (MTPA) and flux-weakening control strategies,
and applies a novel approach that smoothly transitions from one to the other, to ensure differentiability as needed for large-scale (gradient-based) MDO.
Cheng et al.~\cite{cheng2023differentiable} describe this method in more detail.

\subsection{Battery Model}
A simple battery model is implemented that is based on the total amount of energy available in the batteries, which is computed from the battery mass and the battery energy density---assumed to be 400Wh/kg according to Silva et al.~\cite{silva2018vtol}.
The total energy consumption during the aircraft mission is obtained from the power profile, which is computed from the motor model and the time spent in each mission segment. 
Based on the total energy available and consumed, the final state of charge is computed, which is constrained to be 20\% at the end of the mission. 
In future investigations, the battery model will be replaced with an equivalent circuit model that takes into account aspects of the underlying electro-chemistry and time dependent performance degradation. 

    \section{Results}
In this section, the results of an aeroacoustic subsystem optimization study are presented first.
Second, the results from a large-scale MDO study that involves all disciplines are shown, where vehicle gross weight is the objective function subject to multiple constraints. 
Third, parameter sweeps and trade studies are presented, showing the effect of certain constraints and illustrating tradeoffs between opposing optimization objectives.

\subsection{Aeroacoustic Analysis and Optimization} 
\begin{figure}[ht] \begin{center}
\includegraphics[scale=0.92]{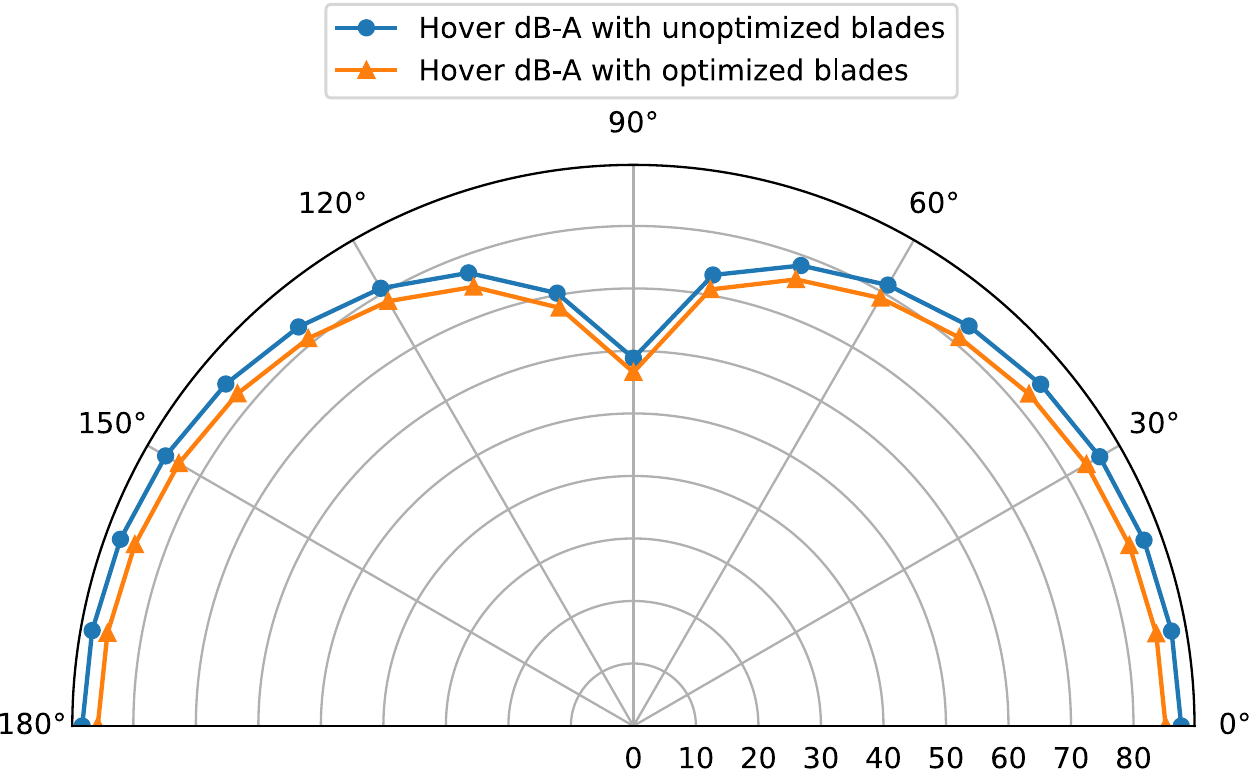}
\caption{Directivity plot of Lift+Cruise concept at an observer distance of 250 ft (76 m).}\label{fig:Directivity}
\end{center}\end{figure}

This section explores the noise signature of the Lift+Cruise configuration in hover, and the results of blade shape optimization to minimize noise.
The noise of UAM vehicles is a significant concern, especially in residential communities.
For this paper, an observer location of 250 ft away from the vehicle was chosen, a benchmark value adapted from an Uber Elevate white paper~\cite{Uber}.
A maximum noise threshold of 67 dB-A is suggested based on existing regulations.
In Fig.~\ref{fig:Directivity}, it can be seen that the noise model does not predict the sound pressure level uniformly, depending on the observer location. 
At 90$^{\circ}$, the observer is located directly below the aircraft and it can be seen that the prediction is significantly lower than when the observer is at a relative angle to the aircraft.
This is due to the previously mentioned limitation of the tonal noise model.
The plot highlights the importance of considering noise in large-scale analyses as the constraint of 67 dB-A is active at this altitude. 

\begin{figure}[ht] \begin{center}
\includegraphics[scale=0.9]{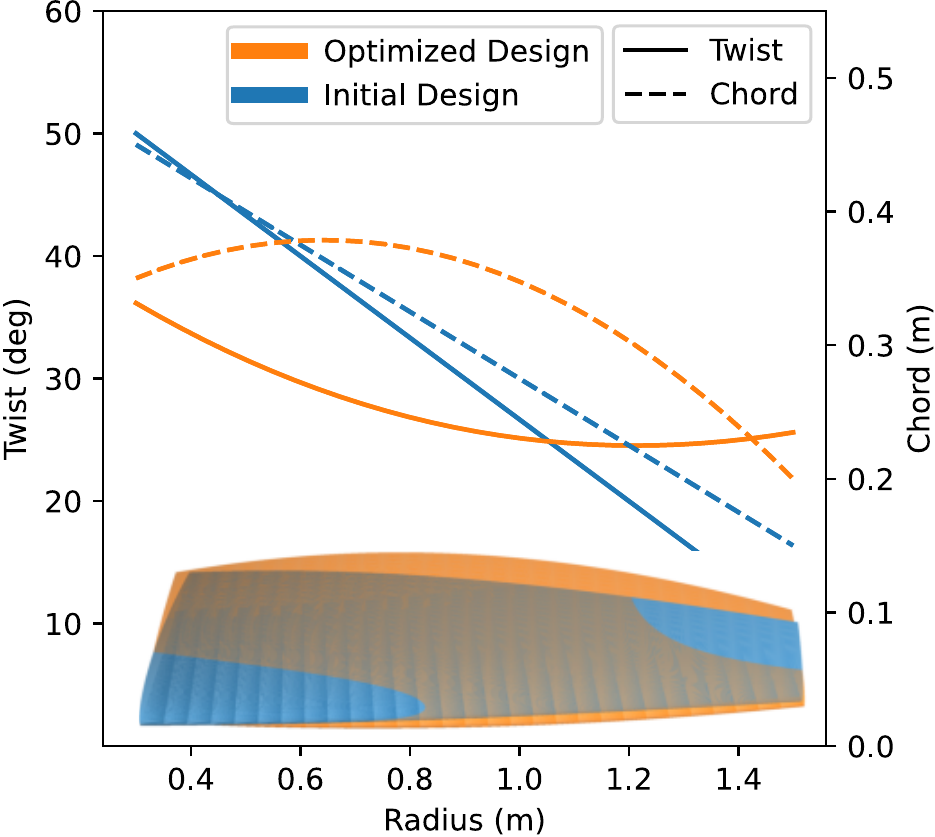}
\caption{Blade shape optimization to minimize noise only.}\label{fig:blade shape opt}
\end{center}\end{figure}
Optimization of the blade shape decreases the overall noise footprint, shown by the orange line in Fig.~\ref{fig:Directivity}.
The MDO problem is given in Tab.~\ref{tab:opt_prob_blades}, showing the objective, design variables, and constraints. 
The objective is a weighted average of the total sound pressure level ($\,\textrm{SPL}_{\textrm{total}}$), given in dB-A, and the mean figure of merit of the eight lift rotors ($\,\overline{\textrm{FOM}}$), where the weighting factor $w$ varies between zero and one. 
For the results shown in Fig.~\ref{fig:Directivity} and Fig.~\ref{fig:blade shape opt}, $w=1$, meaning the objective is to minimize noise only.
The only constraint for this problem is that the aircraft needs to be trimmed while hovering. 
Other mission segments are not considered.

\begin{table}[!htb]
\begin{center}
\caption{Aeroacoustic Optimization Problem}
\begin{tabular}{c l c}
\hline
 & \textbf{Variable/function} & \textbf{Quantity} \\ 
 \hline
 \hline
minimize  
& $w \cdot\,\textrm{SPL}_{\textrm{total}} - (1-w) \cdot\,\overline{\textrm{FOM}}$  &         \\
w.r.t.
& blade twist (4$\times$8) & 32\\
& blade chord (3$\times$8) & 24\\
& lift rotor speed & 8\\
& \emph{\# design variables} & 64 \\ \hline
subject to 
& trim residual norm & 1\\
& \emph{\# constraints} & 1 \\ \hline
\end{tabular}
\label{tab:opt_prob_blades}
\hspace{\fill}
\end{center}
\end{table}


The blade shape is discretized with three B-spline control points for the chord distribution and four for the twist distribution for each of the eight lift rotors, whose radii are kept constant. 
These rotors have two blades based on the baseline Lift+Cruise configuration.
Initially, both profiles are assumed to vary linearly. 
The initial and optimized blade shapes are shown more closely in Fig.~\ref{fig:blade shape opt}. 
An interesting feature of the ``quiet" blade (shown in orange) is that the chord distribution increases away from the hub before decreasing toward the tip. 
It should be noted that the thickness component of the tonal noise is not minimized since the BEM model uses a single airfoil along the span. 

\begin{figure}[ht] \begin{center}
\includegraphics[scale=1]{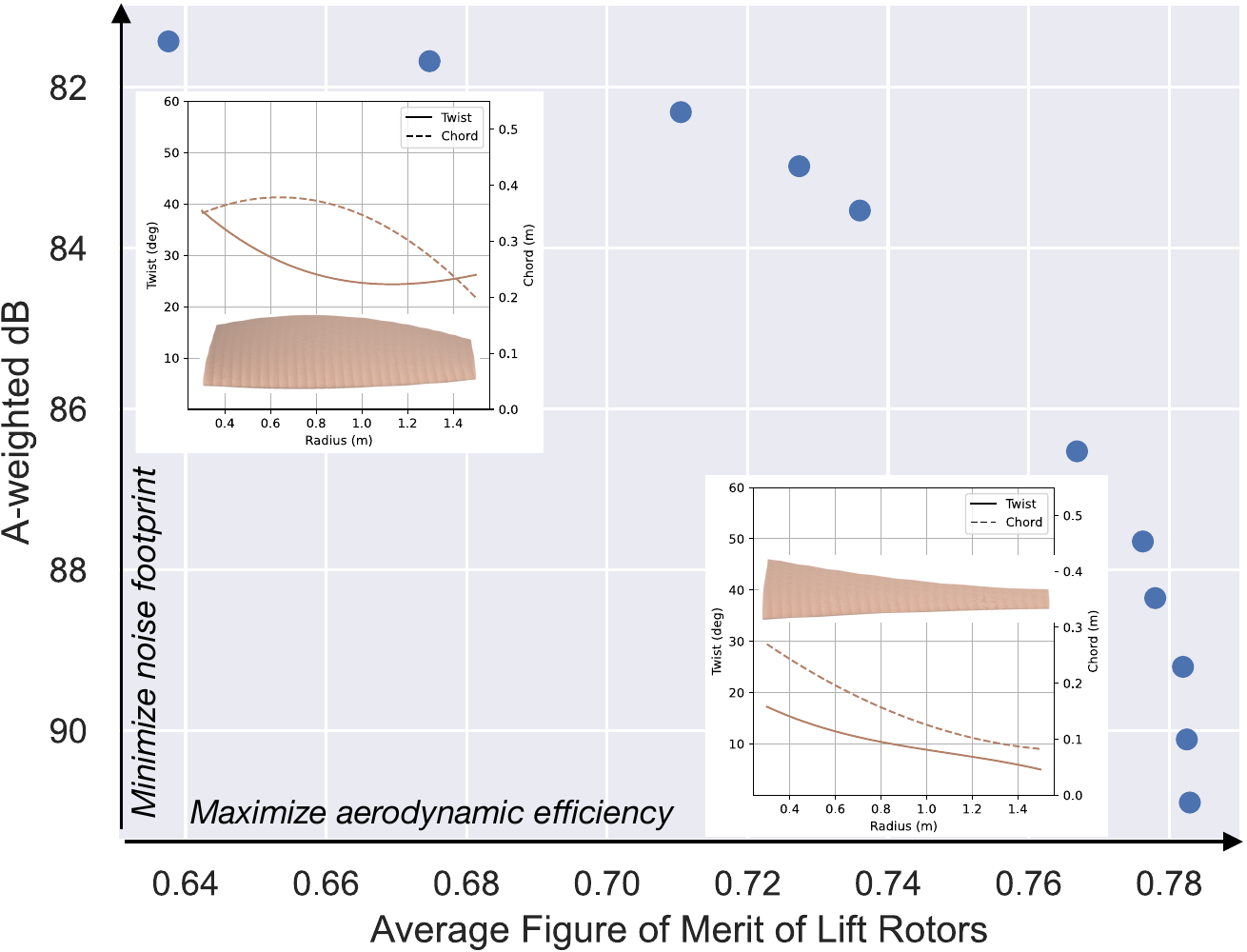}
\caption{Pareto front showing the trade off between minimizing noise and maximizing rotor aerodynamic efficiency.}\label{fig:noise_fom_pareto}
\end{center}\end{figure}

Lastly, the trade off between minimizing rotor noise and maximizing aerodynamic efficiency is shown in Fig.~\ref{fig:noise_fom_pareto} in the form of a Pareto front.
Here, the weighting factor $w$ is varied from zero to one in increments of 0.1, and the observer is at a distance of 250 ft and an angle of 45 degrees.
The design variables and constraints are the same as before.
The Pareto front shows that maximizing aerodynamic efficiency comes at the expense of a louder blade design and vice versa, confirming a known trend.
For the most aerodynamically efficient blade with a figure of merit of just over 0.78, the total noise is roughly 91 dB-A whereas the the quietest blade emits about 81 dB-A with a figure of merit of just below 0.64.
This trade off is an important consideration since high aerodynamic efficiency is desirable in eVTOL design to offset low battery energy densities. 
At the same time, strict regulations are in place to protect communities from excessive noise, limiting the ability to optimize for efficiency only. 
In terms of the blade shape, which is plotted for the extreme cases of the Pareto front, the results agree with what is expected qualitatively as aerodynamically efficient designs are more slender and have a lower twist distribution (in hover).

\subsection{Large-Scale MDO}
The statement of the large-scale MDO problem is given in Tab.~\ref{tab:opt_prob}, broken down into design variables and constraints.
For this study, the objective is chosen as gross weight, which is to be minimized with respect to 109 design variables while satisfying 17 constraints.
Most design variables are of geometric nature and primarily affect the structural mass estimates. 
For the blade shape parameterization, only two B-spline control points were chosen for the chord distribution, while four were maintained for twist. 
Operational variables such as rotor speed or aircraft orientation (e.g., pitch angle), are used to trim the aircraft in every mission segment. 
In terms of constraints, the trim residual is an equality constraint set to zero such that forces and moments are balanced.
The final state of charge is an inequality constraint with a lower bound of 0.2. 
The sound pressure level is also an inequality constraint with an upper bound of 67 dB-A at an altitude of 250 ft (recall Uber Elevate white paper).
The noise of the pusher rotor was not constrained due to the relatively large distance to the ground. 
It was found that at the cruising altitude of roughly 3300 ft, the total noise of the pusher rotor was about 45 dB-A.
Since the cruise altitude is a design variable, a constraint is needed to ensure that the altitude at the end of climb matches that of cruise. 
For motor sizing, a maximum motor torque constraint is set for the pusher motor, which is sized by its diameter and length. 
This constraint ensures that the motor is sized sufficiently large in order to provide enough torque during the climb and cruise segment. 
Similar constraints for the lift motors are not set.
This is because the motor model is more computationally expensive than other models and adding eight additional constraints of this kind slows down optimization time significantly.  
It is ensured that the lower bound of the lift motor weights is high enough such that the torque constraint is not active. 
The stall speed constraint ensures a buffer of at least 5 m/s between the operational and the stall speed. 
The rest of the constraints ensure that rotors do not overlap and that the symmetry is maintained for motor weights and rotor radii. 
Future, more advanced versions of CADDEE will automate symmetry, such that these kind of constraints will not need to be made. 

\begin{table}[!hbt]
\begin{center}
\caption{Large-Scale Optimization Problem}
\begin{tabular}{c l c} 
\hline
 & \textbf{Variable/function} & \textbf{Quantity} \\ 
\hline
\hline
minimize  
& gross weight   &         \\
w.r.t.
& rotor radii ($\times$9) & 9 \\
& blade twist (4$\times$9) & 36\\
& blade chord (2$\times$9) & 18\\
& motor length ($\times$9) & 9\\
& motor diameter ($\times$9) & 9\\
& wing area, wing aspect ratio & 2\\
& wing twist & 5\\
& HT area, aspect ratio, location & 3\\
& battery mass, battery location & 2\\
& lift rotor speed & 8\\
& propeller speed ($\times$2) & 2\\
& vehicle pitch angle ($\times$2) & 2\\
& horizontal tail deflection ($\times$2) & 2\\
& cruise altitude & 1\\
& end-of-climb altitude & 1 \\
\hline
& \emph{\# design variables} & 109 \\ \hline
subject to 
& trim residual norm & 1\\
& final state of charge & 1\\
& stall speed & 1\\
& climb-cruise altitude continuity & 1 \\
& sound pressure level & 1\\
& rotor symmetry & 4\\
& rotor tip clearance & 2 \\
& maximum motor torque ($\times$2) & 2\\
& motor symmetry & 4\\

& \emph{\# constraints} & 17 \\ 
\hline
\end{tabular}
\label{tab:opt_prob}
\hspace{\fill}
\end{center}
\end{table}

\begin{figure}[!hbt] \begin{center}
\includegraphics[scale=1.2]{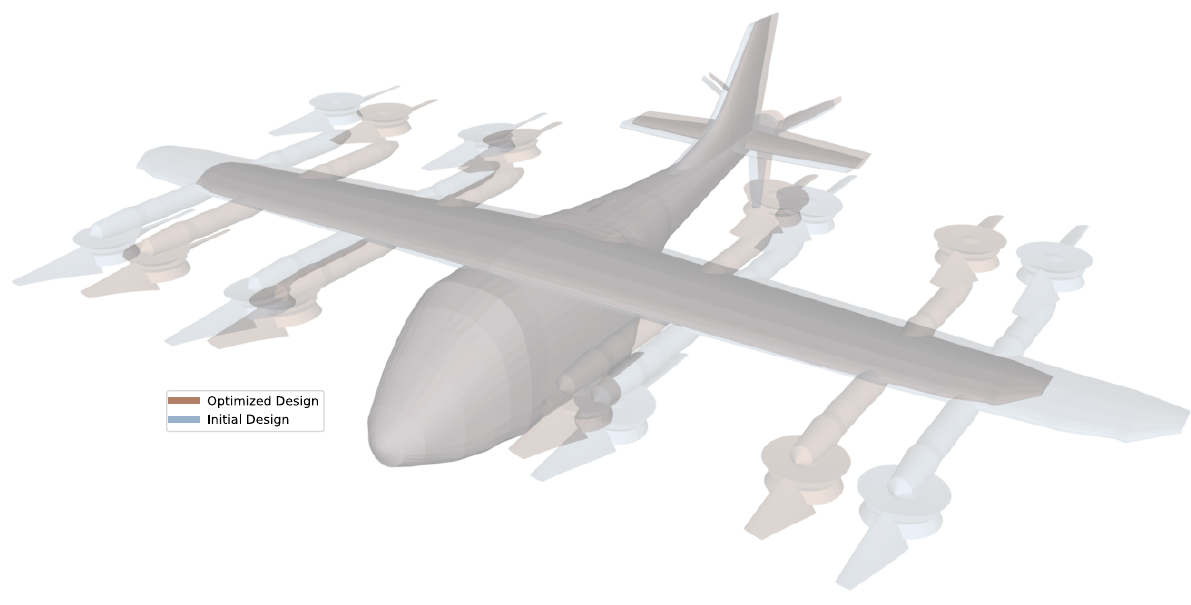}
\caption{Initial and optimized geometry}\label{fig:optimized geometry}
\end{center}\end{figure}

\begin{figure}[ht] \begin{center}
\includegraphics[scale=1]{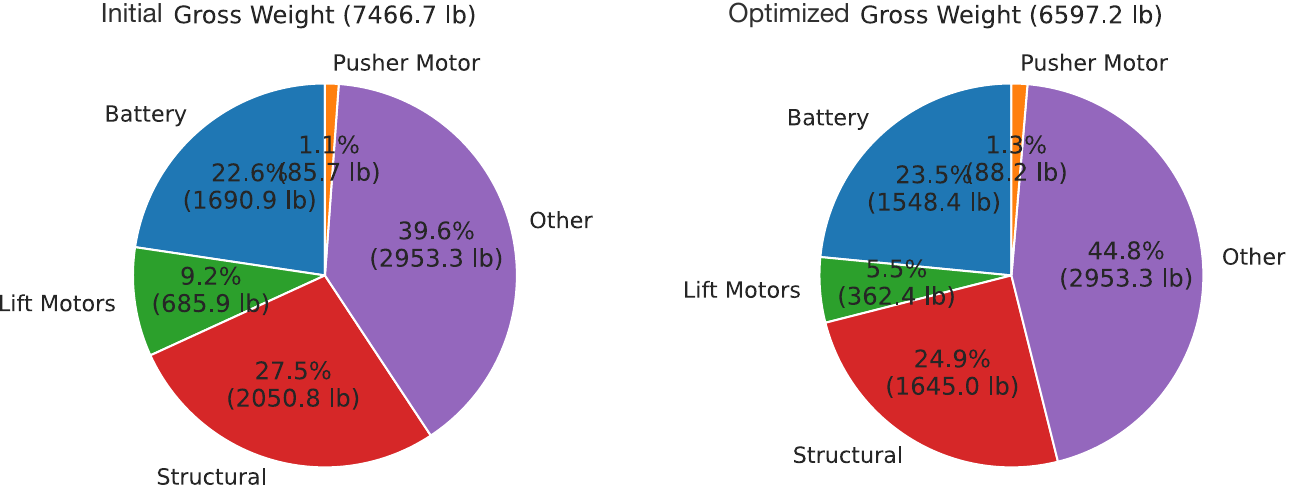}
\caption{Initial and optimized gross weight}\label{fig:weight pie charts}
\end{center}\end{figure}

The initial and optimized aircraft geometry is shown in Fig.~\ref{fig:optimized geometry}.
It can be seen that the wing span has been significantly reduced in order to minimize the structural weight. 
The horizontal and vertical tail are only slightly reduced in size. 
A more detailed breakdown of gross weight before and after optimization is shown in Fig.~\ref{fig:weight pie charts}.  
After optimization, the gross weight is reduced by 11.4\% from approximately 7500 lb to just under 6600 lb. 
The bulk of the weight reduction is achieved by reducing the battery and structural weight. 
In addition, the lift motors are also downsized to save additional weight. 
Only the pusher motor weight is increased by about 4 pounds, in order to meet the maximum torque constraint. 
It should be noted that the non-structural mass, which includes estimates for passengers, avionics, landing gear, wiring etc., remains constant due to a lack of an accurate parametric model to predict these quantities.

\begin{figure}[!hbt] \begin{center}
\includegraphics[scale=1.5]{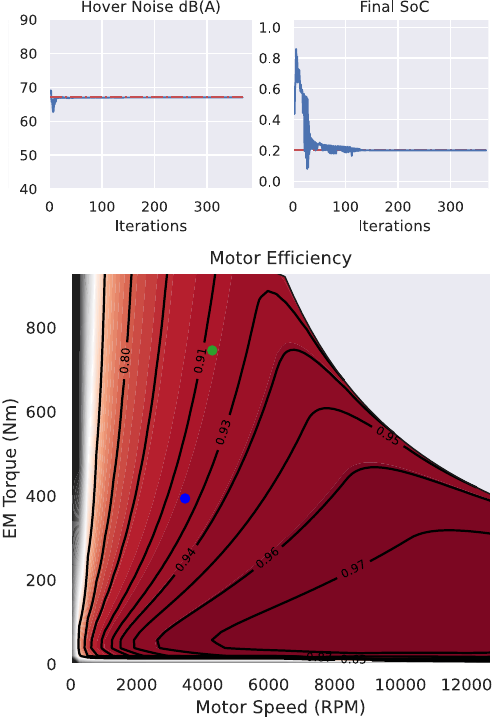}
\caption{Final state-of-charge constraint, noise constraint, and motor efficiency map. Cruise and climb are shown in blue and green, respectively.}\label{fig:optimization constraints}
\end{center}\end{figure}

Next, two optimization constraints are shown in Fig.~\ref{fig:optimization constraints} as a function of the optimization iteration. 
The noise and final state-of-charge constraints are both active, which means that they play a role in the aircraft sizing.
At the end of the optimization, both constraints are met.
It should be noted that according to the models used in this study, the noise constraint cannot be met if enforced at every observer angle for a distance of 250 ft away from the aircraft. Therefore, it is enforced only at 85 degrees to generate the results presented in this section.  
The 85-degree observer location is chosen because the tonal noise model grossly under-predicts at 90 degrees (see Fig.~\ref{fig:Directivity}). Thus, a noise constraint enforced at 85 degrees approximates the true noise constraint with the observer directly below the aircraft, but with enough distance from 90 degrees such that the region of large error in the model is avoided.

In addition, a motor efficiency map shows the motor efficiency as a function of the torque and rotational speed. 
As expected, the motor efficiency is high for the majority of the operating regime. 
The efficiency map also shows that the points of operation of the pusher motor during the climb (green) and cruise (blue) segment. 
It can be seen that during climb, the torque requirement is more stringent, causing the motor to operate closer to its physical limit. 
This means that the climb segment primarily sizes the motor, which is expected. 

\begin{figure}[!hbt] \begin{center}
\includegraphics[scale=1]{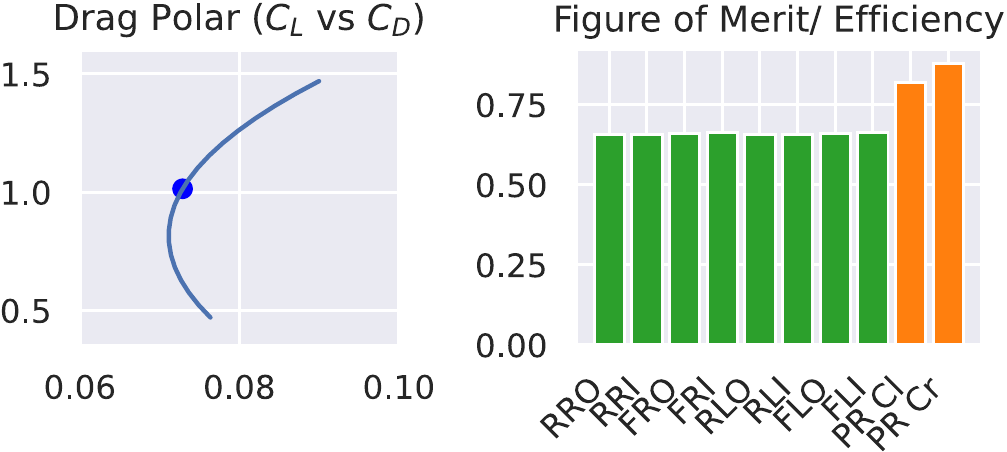}
\caption{
    Drag polar and rotor aerodynamic efficiency. 
    F: front;
    R: rear/right;
    L: left;
    O: outer;
    I: inner;
    Cl: climb;
    Cr: cruise.
}\label{fig:aero efficiency}
\end{center}\end{figure}

Lastly, the aerodynamic efficiency of the optimized design is shown in Fig.~\ref{fig:aero efficiency}.
The aircraft drag polar shows the operational lift-to-drag ratio in cruise (blue dot), which is about 13.5, a reasonable value for the Lift+Cruise configuration. 
It should be noted that the minimum drag of about 0.07 seems unreasonably high. 
A reason for this could be a lack of accuracy in the drag area estimates that were used for components whose drag is not computed by the aerodynamic solver (VLM), such as the fuselage, rotor hubs, booms, vertical tail and landing gear. 

In terms of the rotor-aerodynamic efficiency, the average figure of merit for the hover rotors (green) is about 0.65. 
While this value seems low, it is to be expected based on the results from the aeroacoustic optimizations, where it was shown that aerodynamic efficiency comes the expense of louder rotors. 
In this case, the active noise constraint causes an aerodynamic penalty in order to be satisfied, resulting in a relatively low figure of merit. 
For the pusher rotor, the efficiency in climb (denoted PR CL) is slightly lower than in cruise, which is expected. 
The cruise efficiency is relatively high, which makes sense due to the active final state-of-charge constraint.
Since the cruise segment(s) make up the majority of the mission profile, most of the energy expenditure is expected during these stages of the mission.
In order to satisfy the final state-of-charge constraint, the pusher rotor needs to be as efficient as possible, especially when the battery weight is being saved.

\subsection{Parameter Sweeps and Trade Studies}
Finally, a set of large-scale MDO studies including the same design variables and constraints as shown in Tab.~\ref{tab:opt_prob} are presented to further explore the design space by varying constraining parameters (e.g., battery energy density) and evaluating the effect of opposing optimization objectives.

\begin{figure}[!hbt] \begin{center}
\includegraphics[scale=0.9]{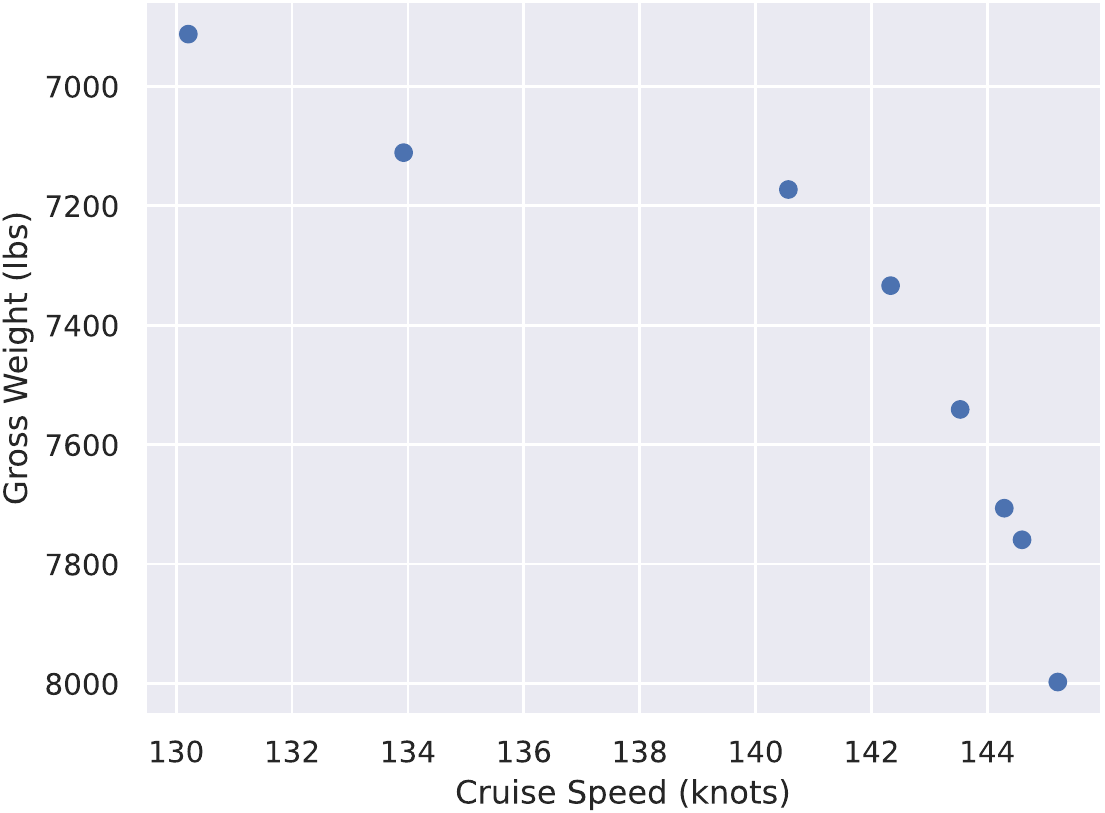}
\caption{Pareto front showing the trade off between minimizing gross weight and maximizing cruise speed}\label{fig:weight_speed_pareto}
\end{center}\end{figure}

In Fig.~\ref{fig:weight_speed_pareto}, a Pareto front is shown, representing a mixed objective function of gross weight and cruise speed. 
These objectives have opposite effects since in order to fly faster, the aircraft needs to overcome more aerodynamic drag and hence requires a larger battery. 
Since drag has a quadratic dependence on speed, this effect should become more pronounced at higher speeds.
This is in fact the case as the gross weight increases rapidly at higher cruise speeds by over 1000 lb when compared at the ends of the Pareto front, corresponding to an increase of over 10\%. 
 For UAM vehicles, gross weight can be viewed as a proxy for acquisition cost, and therefore, a lighter aircraft is desirable.
 At the same time, faster cruise speeds decrease operational time, which allows for a higher volume of passenger trips. 
 Of course, other factors like cost of energy also need to be considered when evaluating the best trade off. 

\begin{figure}[!hbt] \begin{center}
\includegraphics[scale=0.9]{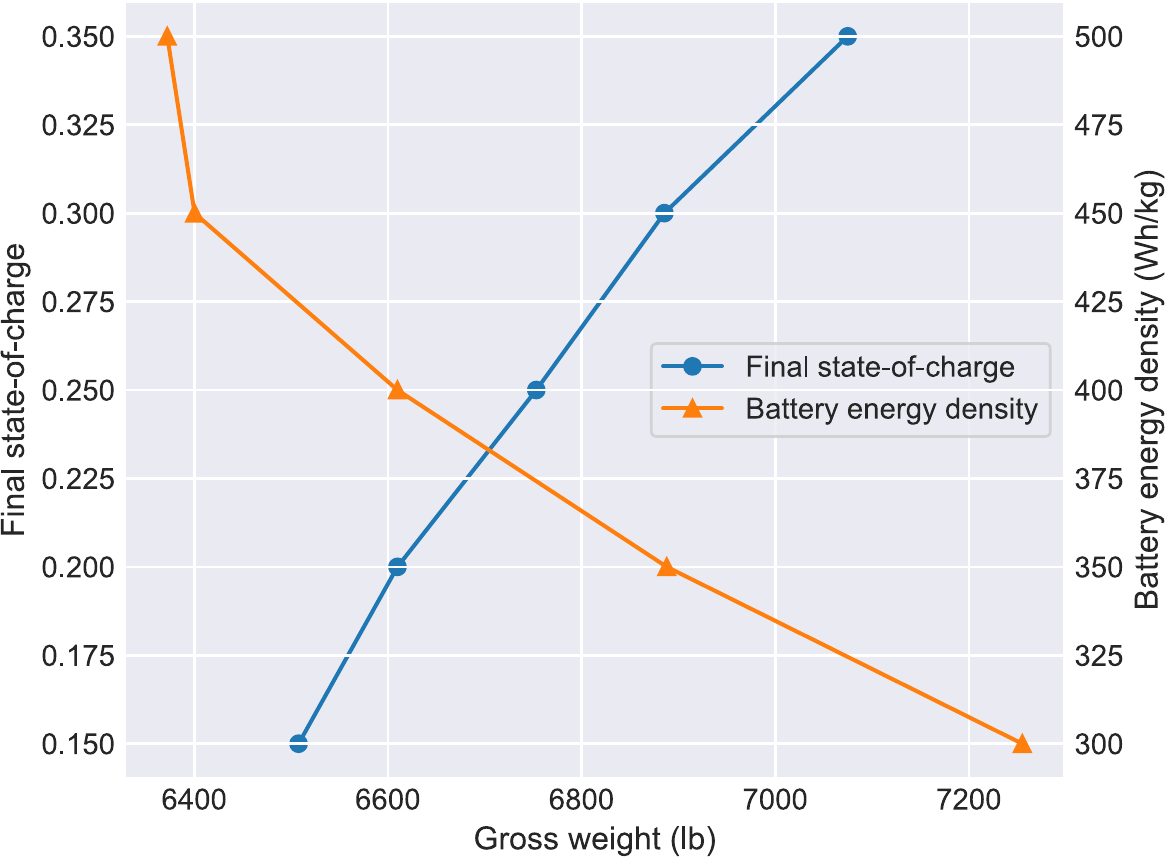}
\caption{Effect of battery energy density and final state-of-charge constraint on the optimized vehicle gross weight.}\label{fig:parameter_sweeps}
\end{center}\end{figure}

Figure~\ref{fig:parameter_sweeps} shows the results of performing the the same large-scale MDO problem as in Tab.~\ref{tab:opt_prob} for different values of battery energy density  and the final state-of-charge constraint.
These quantities were changed independently while keeping the other quantity at the baseline value.
The results agree qualitatively with what is expected and confirm the importance of high energy densities for electrified aviation.
 
    \section{Conclusions}
This paper presents the results of a set of large-scale MDO studies of a NASA eVTOL concept (Lift+Cruise). 
These results were generated using a new design framework still under development, called the Comprehensive Aircraft high-Dimensional DEsign Environment (CADDEE). 
The new aircraft design framework enables geometry-centric analysis and design and automates derivative computation for large-scale MDO.
The most important results of this effort are summarized as follows:
\begin{enumerate}
    \item With the CADDEE framework, the run time for large-scale MDO studies is less than one hour on a standard desktop computer.
    \item The application of large-scale MDO to the Lift+Cruise concept results in a 11.4\% reduction of vehicle gross weight with 109 design variables and 17 constraints across disciplines including aerodynamics, propulsion, structural sizing, acoustics, stability, motor, and battery analysis.
    \item Several active constraints are successfully enforced, addressing the vehicle's noise footprint, stability (trim-state), motor sizing, final state-of-charge, and others. 
    \item A Pareto front of noise in hover versus figure of merit confirms that maximizing rotor aerodynamic efficiency comes at the expense of a larger noise footprint. It was found that for a maximum average figure of merit of 0.78, the total noise is about 91 dB-A whereas for a minimum total noise of 81 dB-A the average figure of merit drops to below 0.64.
    \item A Pareto front of maximum cruise speed versus vehicle gross weight confirms that cruise speed has a significant effect on vehicle gross weight. According to these results, an 11\% increase in cruise speed leads to a 15\% increase in gross weight.
\end{enumerate}
    \section*{Acknowledgments}

The authors would like to thank PhD student Mark Sperry for his help generating the visualizations presented in this paper and PhD student Luca Scotzniovsky for his expertise on motor modeling. 

The material presented in this paper is supported by NASA under award No. 80NSSC21M0070.

    \section*{References}
    \printbibliography[heading=none]

\end{refsection}

\end{document}